\documentclass[epj, nopacs]{svjour}

\usepackage[utf8]{inputenc}
\usepackage{graphicx}
\usepackage{units}
\usepackage{color}
\usepackage{bbold}
\usepackage{amsmath}
\usepackage{amssymb}
\usepackage{cite}
\usepackage{comment}

\usepackage[pdftex, colorlinks=true, linkcolor=myblue, citecolor=myblue,
urlcolor=myblue]{hyperref}

\definecolor{myblue}{rgb}{0,0,1}

\newcommand{\q}{\mathbf{q}}
\newcommand{\R}{\mathbf{R}}

\begin{document}

\title{Tunable plasmon polaritons in arrays of interacting metallic nanoparticles}

\author{Guillaume Weick\inst{1} \and Eros Mariani\inst{2}}
\institute{
Institut de Physique et Chimie des Mat\'{e}riaux de Strasbourg, Universit\'{e} de
Strasbourg, CNRS UMR 7504, F-67034 Strasbourg, France
\and
Centre for Graphene Science, Department of Physics and Astronomy, University of Exeter, EX4 4QL Exeter, United Kingdom}

\abstract{
We consider a simple cubic array of metallic nanoparticles supporting extended
collective plasmons that arise from the near-field dipolar interaction between
localized surface plasmons in each nanoparticle. We develop a fully analytical
quantum theory of the strong-coupling regime between these collective plasmons and
photons resulting in plasmon polaritons in the nanoparticle array.
Remarkably, we show that the polaritonic band gap and the dielectric function
of the metamaterial can be significantly modulated by
the polarization of light. We unveil how such an anisotropic behavior in the
plasmonic metamaterial is crucially mediated by the dipolar
interactions between the nanoparticles despite the symmetry of the underlying lattice.
Our results thus pave the way towards the realization of tunable quantum plasmonic 
metamaterials presenting interaction-driven birefringence.
%
%\PACS{
%      {73.20.Mf}{Collective excitations (including excitons, polarons, plasmons and other charge-density excitations)}   \and
%      {71.36.+c}{Polaritons (including photon-phonon and photon-magnon interactions)} \and
%      {78.67.Bf}{Nanocrystals, nanoparticles, and nanoclusters} \and
%      {73.22.Lp}{Collective excitations}
%     } 
}

\maketitle

%===========================================================================
%===========================================================================
%===========================================================================
%===========================================================================
\section{Introduction}
The ability to manipulate light at
subwavelength scales beyond the diffraction limit of traditional optics is at the very heart of the present research in 
plasmonic metamaterials \cite{barne03_Nature, maier} and transformation optics
\cite{pendry12_Science}. 
Metamaterials have already been shown to exhibit exciting new properties such as negative refractive
index \cite{vesel68_SPU, smith00_PRL, shelb01_Science}, 
perfect lensing \cite{pendr00_PRL, fang05_Science}, electromagnetic
invisibility cloaking \cite{leonh06_Science, pendr06_Science, schur06_Science}, ``trapped
rainbow" slow light \cite{tsakm07_Nature}, and the ability to perform
mathematical operations (``metamaterial analog computing")
\cite{silva14_Science}. In this paper we explore the possibility to design novel
quantum plasmonic metamaterials \cite{tame13_NaturePhys} with a highly tunable optical response based on arrays of interacting metallic nanoparticles.

While isolated metallic nanoparticles have been already successfully exploited 
to confine electromagnetic radiation down to the nanometer scale, the focus has now 
shifted to the exploration of ordered plasmonic arrays of metallic
nanoparticles. In these systems the interactions between the nanoparticles lead to
dramatic changes in the collective plasmonic properties of the array as compared to those 
of the individual nanoparticles.  This opens up new perspectives for confining and guiding light at
subwavelength scales, as has been explored experimentally in one- and
two-dimensional arrays of gold \cite{krenn99_PRL, maier02_PRB, felid02_PRB} and
silver nanoparticles \cite{maier03_NatureMat, sweat05_PRB, polyu11_NL} and
studied theoretically by means of classical electromagnetic calculations
\cite{quint98_OL, brong00_PRB,park04_PRB, weber04_PRB, koend06_PRB, koend09_NL}.
In this context, we have recently exploited the tunability of 
the near-field interaction in a two-dimensional honeycomb array of metallic
nanoparticles supporting 
Dirac-like collective plasmons \cite{weick13_PRL}.

The optical response of plasmonic metamaterials is determined by their dielectric 
function, which results in the reflection and transmission coefficients. 
In order to calculate the dielectric function it is crucial to identify the
eigenmodes responsible for transporting electromagnetic radiation in arrays of
metallic nanoparticles. 
These modes, stemming from the coupling between light and plasmons, have been
extensively investigated in one- and two-dimensional nanoparticle arrays 
in the weak coupling regime
\cite{krenn99_PRL, maier02_PRB, felid02_PRB, maier03_NatureMat, sweat05_PRB,
polyu11_NL, quint98_OL, brong00_PRB,park04_PRB, weber04_PRB, koend06_PRB,
koend09_NL}, where the semiclassical theory of radiation provides a
satisfactory description of the optical properties.
In contrast, it is well established \cite{fano56_PR, hopfi58_PR} that this semiclassical 
picture is inadequate for studying the fundamental nature of absorption processes in periodic systems
that require a full quantum treatment of the strong coupling regime between light and
matter, 
taking into account the conservation of crystal momentum between photons and
polarization fields. This treatment gives rise to
new quasiparticles --- termed polaritons ---   
that were first studied in the late 1950's by Fano \cite{fano56_PR} and Hopfield \cite{hopfi58_PR} in the
context of excitons in bulk solids.
Polaritons are coherent superpositions of light and matter quantum fields and
represent the natural quasiparticles involved in absorption processes in
periodic systems. In the light of this analysis, 
the exploration of three-dimensional arrays of interacting metallic
nanoparticles thus requires a full quantum analysis of the strong coupling
regime between photons and collective plasmons that is expected to give rise to
\textit{plasmon polaritons}. 
This is the purpose of the present theoretical paper. 

As a proof of concept, we hence explore plasmon polaritons in a simple cubic array of metallic
nanoparticles. By means of a fully quantum-mechanical approach, 
we analytically unveil the plasmon polariton band structure,  
modeling the localized surface plasmon in each nanoparticle as a point
dipole interacting with the neighboring ones through a near-field interaction.
Such an interaction results in a plasmon polariton band structure that is highly tunable
with the polarization of light, giving rise to 
dramatic effects
which would otherwise be absent in noninteracting systems.
Remarkably, we show that the 
plasmon polariton band gap can be tuned by about $\unit[50]{\%}$. 
Our prediction should
thus be clearly observable in the frequency- and wavevector-dependent dielectric function of the
metamaterial, resulting in an \textit{interaction-driven birefringence} despite
the symmetric lattice structure of the array. Our analytical treatment can be
easily generalized to other three-dimensional lattices that are thus expected to
exhibit a similar tunable optical response. Our theoretical proposal could be
experimentally realized in self-organized arrays of metallic nanoparticles 
capped with molecular linkers such as thiol chains \cite{donni07_AdvMater} and DNA 
\cite{tan11_NatureNanotech}. 

Our paper is organized as follows: in Sect.\ \ref{sec:arrays} we present our
model of localized surface plasmons in a simple cubic array of interacting metallic nanoparticles, while
Sect.\ \ref{sec:plasmon} presents our results for the collective plasmon
dispersion. In Sect.\ \ref{sec:light} we discuss the coupling of these
collective modes to light, and deduce the plasmon polariton dispersion in
Sect.\ \ref{sec:PP}. We present our conclusions in Sect.\
\ref{sec:ccl}.

%===========================================================================
%===========================================================================
%===========================================================================
%===========================================================================
\section{Arrays of interacting metallic nanoparticles}
\label{sec:arrays}
We consider an ensemble of identical spherical metallic
nanoparticles of radius $r$, each containing $N_\mathrm{e}$ valence electrons, and
forming a simple cubic lattice with $\mathcal{N}\gg1$ lattice sites and lattice constant
$a$.\footnote{In the remainder of the paper, we consider Born-von Karman periodic boundary conditions.} Each nanoparticle supports a localized surface plasmon which corresponds to a collective 
excitation of the electronic center of mass that can be modeled as a point
dipole oscillating at the Mie frequency
$\omega_0$ \cite{kreibig}. 
This is justified as long as the size of the nanoparticle is much smaller than the 
wavelength associated with light at frequency $\omega_0$.\footnote{As detailed
in Sect.\ \ref{sec:PP}, 
we consider nanoparticles having a localized surface plasmon resonance in 
the visible range of the spectrum and with typical size of the order of a 
few nanometers.}
For a nanoparticle in vacuum, the Mie frequency takes the simple form
$\omega_0=(N_\mathrm{e}e^2/4\pi\epsilon_0m_\mathrm{e}r^3)^{1/2}$, 
where  $-e$ and $m_\mathrm{e}$ are the electron charge and mass, respectively, and where
$\epsilon_0$ is the vacuum permittivity.
The noninteracting part of the Hamiltonian describing the independent localized
surface plasmons on the cubic lattice
sites hence reads \cite{gerch02_PRA, weick06_PRB}
\begin{equation}
\label{eq:H_0}
H_0=\sum_{\mathbf{R}}\left[\frac{\Pi^2(\mathbf{R})}{2M}+\frac{M}{2}\omega_0^2h^2(\mathbf{R})\right],
\end{equation}
where $h(\mathbf{R})$ denotes the electronic center-of-mass displacement corresponding to a
nanoparticle located at position $\mathbf{R}$, $\Pi(\mathbf{R})$ is the conjugated momentum
to $h(\mathbf{R})$ and
$M=N_\mathrm{e}m_\mathrm{e}$ is the total electronic mass per nanoparticle.
The point dipole corresponding to each localized surface plasmon has thus a dipole 
moment $\mathbf{p}=-Qh(\mathbf{R})\hat{\mathbf{p}}$, with 
$\hat{\mathbf{p}}$ the unit vector indicating its direction and $Q=N_\mathrm{e}e$
the total electronic charge.

Assuming that the wavelength associated with the resonance frequency of each
localized surface plasmon is much larger than the interparticle distance
$a$, retardation effects can be ignored and the interparticle coupling occurs
via quasistatic near-field interaction.
Moreover, when $a\gtrsim 3r$, the latter can be modeled \cite{brong00_PRB, park04_PRB} as
a coupling between two point dipoles 
$\mathbf{p}$ and $\mathbf{p}'$ located at
$\mathbf{R}$ and $\mathbf{R}'$, respectively, with interaction potential 
\begin{equation}
\label{Vdip}
V_\mathrm{dip}=
\frac{\mathbf{p}\cdot\mathbf{p}'
-3(\mathbf{p}\cdot\mathbf{n})
(\mathbf{p}'\cdot\mathbf{n})}
{4\pi\epsilon_0|\mathbf{R}-\mathbf{R}'|^3}, 
\end{equation}
where $\mathbf{n}=(\mathbf{R}-\mathbf{R}')/{|\mathbf{R}-\mathbf{R}'|}$.
In what follows, we impose that, due to the electric field associated with
light, all localized surface plasmons
are polarized in the same direction
$\hat{\mathbf{p}}=\sin{\theta}\cos{\varphi}\,\hat{\mathbf{x}}+\sin{\theta}\sin{\varphi}\,\hat{\mathbf{y}}+\cos{\theta}\,\hat{\mathbf{z}}$, 
where $\theta$ is the angle between $\hat{\mathbf{p}}$ and
$\hat{\mathbf{z}}$, and $\varphi$ the angle between the projection of
$\hat{\mathbf{p}}$ in the $xy$ plane and $\hat{\mathbf{x}}$.
This is justified by the absence of retardation effects in our point-like dipole
model. From \eqref{Vdip} the interaction Hamiltonian between localized surface plasmons thus reads
\begin{equation}
\label{eq:H_int}
H_\mathrm{int}=\frac{Q^2}{8\pi\epsilon_0a^3}\sum_{\mathbf{R}}\sum_{j=1}^3
\mathcal{C}_jh(\mathbf{R})\big[h(\mathbf{R}+\mathbf{e}_j)+h(\mathbf{R}-\mathbf{e}_j)\big],
\end{equation}
with
\begin{equation}
\mathcal{C}_j=1-3\left[\sin^2{\theta}\left(\delta_{j1}\cos^2{\varphi}+\delta_{j2}\sin^2{\varphi}\right)
+\delta_{j3}\cos^2{\theta}\right]
\end{equation}
and where $\mathbf{e}_1=a\,\hat{\mathbf{x}}$, $\mathbf{e}_2=a\,\hat{\mathbf{y}}$ 
and $\mathbf{e}_3=a\,\hat{\mathbf{z}}$. Only interactions between
nearest neighbors are taken into account in the Hamiltonian \eqref{eq:H_int}
since, as detailed in Appendix \ref{sec:beyond}, 
the interactions beyond nearest neighbors do not qualitatively change the
collective plasmon dispersion, as is also the case for 
metallic nanoparticle arrays with other geometries \cite{brong00_PRB, weick13_PRL}.

%===========================================================================
%===========================================================================
%===========================================================================
%===========================================================================
\section{Collective plasmon dispersion}
\label{sec:plasmon}
Introducing the bosonic operator 
\begin{equation}
\label{eq:b}
b_\mathbf{R}=\sqrt{\frac{M\omega_0}{2\hbar}}h(\mathbf{R})
+\mathrm{i}\frac{\Pi(\mathbf{R})}{\sqrt{2M\hbar\omega_0}}
\end{equation}
which annihilates a localized surface plasmon at lattice site $\R$
and its momentum space representation $b_\mathbf{q}$ through
$b_\mathbf{R}=\mathcal{N}^{-1/2}\sum_{\mathbf{q}}\exp{(\mathrm{i}\mathbf{q}\cdot\mathbf{R})}b_\mathbf{q}$,
the Hamiltonian representing the collective plasmons,
\begin{equation}
\label{eq:H_pl_0}
H_\mathrm{pl}=H_\mathrm{0}+H_\mathrm{int}, 
\end{equation}
with $H_0$ and $H_\mathrm{int}$ defined in 
\eqref{eq:H_0} and \eqref{eq:H_int}, respectively, transforms into 
\begin{equation}
\label{eq:H_pl}
H_\mathrm{pl}=\hbar\sum_{\mathbf{q}}
\left[
(\omega_0+2\Omega f_\q) b_\q^\dagger b_\q^{\phantom{\dagger}}
+\Omega f_\q(b_\q^\dagger b_{-\q}^\dagger+b_{-\q}b_{\q})
\right], 
\end{equation}
with $\Omega=\omega_0(r/a)^3/2\ll\omega_0$ and 
\begin{equation}
\label{eq:f_q}
f_\q=\sum_{j=1}^3\mathcal{C}_j\cos{(\q\cdot\mathbf{e}_j)}.
\end{equation}

\begin{figure}[tb]
\centerline{\includegraphics[width=\columnwidth]{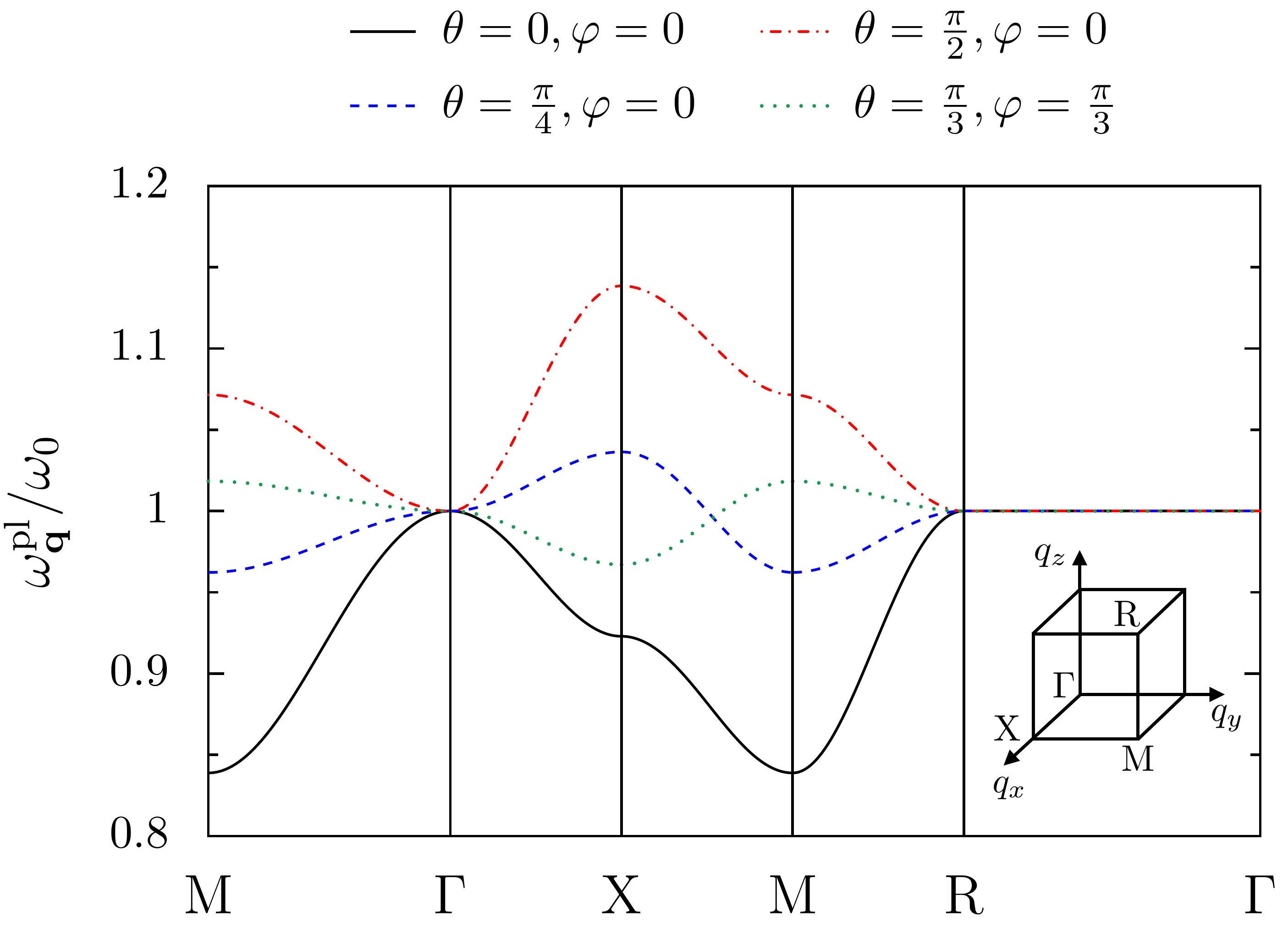}}
\caption{\label{fig:plasmon_dispersion}%
Collective plasmon dispersion relation \eqref{eq:omega_pl} along the
high symmetry axes in the first Brillouin zone, for various
polarization angles $(\theta, \varphi)$. In the figure, $a=3r$.
The inset shows one eighth of the cubic first Brillouin zone together with the high symmetry points.}
\end{figure}

The purely plasmonic problem represented by the Ha\-mi\-lto\-nian \eqref{eq:H_pl} 
can be diagonalized by a Bogoliubov transformation.
The introduction of the new bosonic operators
\begin{equation}
\label{eq:beta}
\beta_\q=\cosh{\vartheta_\q}b_\q-\sinh{\vartheta_\q}b_{-\q}^\dagger, 
\end{equation}
with 
\begin{align}
\cosh\vartheta_\q&=\frac{1}{\sqrt{2}}\left(\frac{1+2\Omega f_\q/\omega_0}{\sqrt{1+4\Omega
f_\q/\omega_0}}+1\right)^{1/2},
\\ 
\sinh\vartheta_\q&=-\frac{\mathrm{sign}(f_\q)}{\sqrt{2}}\left(\frac{1+2\Omega f_\q/\omega_0}{\sqrt{1+4\Omega
f_\q/\omega_0}}-1\right)^{1/2},
\end{align}
leads to 
\begin{equation}
\label{eq:H_pl_diag}
H_\mathrm{pl}=\sum_\q \hbar\omega_\q^\mathrm{pl}\beta_\q^\dagger\beta_\q^{\phantom{\dagger}}, 
\end{equation}
with the collective plasmon dispersion
\begin{equation}
\label{eq:omega_pl}
\omega_\q^\mathrm{pl}=\omega_0\sqrt{1+4\frac{\Omega}{\omega_0}f_\q}.
\end{equation}
It is important to realize that the dispersion (\ref{eq:omega_pl}) can be
tuned by the polarization of the localized surface plasmons which enters the
function $f_\q$ in \eqref{eq:f_q}. This is illustrated
in Fig.\ \ref{fig:plasmon_dispersion} which shows the collective plasmon
dispersion \eqref{eq:omega_pl} along the high symmetry axes in the
first Brillouin zone. As can be seen from the figure, the collective plasmon
dispersion can be dramatically changed by the polarization of the localized surface plasmons.
As will be shown in Sect.\ \ref{sec:PP}, this feature is at the very heart of the tunability of the
polaritonic band gap, of the dielectric function and thus of the resulting
reflection and transmission coefficients of the
metamaterial with the polarization of light.

%===========================================================================
%===========================================================================
%===========================================================================
%===========================================================================
\section{Coupling to light}
\label{sec:light}
The collective plasmons discussed above can be triggered by 
light. As realized by Fano \cite{fano56_PR} and Hopfield \cite{hopfi58_PR},
an adequate description of the strong coupling of elementary excitations to light in
periodic systems can be achieved by
quantizing electromagnetic modes in a cavity that has the same size as the
crystal.
This is a direct consequence of the translational invariance of the nanoparticle
array and the subsequent conservation of photonic and plasmonic crystal momenta. As a
result, energy oscillates back and forth between the two subsystems, such
that the semiclassical view of absorption processes is not appropriate. 
Hence, we describe the photonic modes in the cavity of volume
$\mathcal{V}=\mathcal{N}a^3$ by the Hamiltonian 
\begin{equation}
\label{eq:H_ph}
H_\mathrm{ph}=\sum_\q\hbar\omega_\q^\mathrm{ph}c_\q^\dagger c_\q^{\phantom{\dagger}}, 
\end{equation}
where $c_{\mathbf{q}}$ ($c_\q^\dagger$) annihilates (creates) a photon with momentum
$\mathbf{q}$ and transverse polarization $\hat{\boldsymbol{\epsilon}}$
($\mathbf{q}\cdot\hat{\boldsymbol{\epsilon}}=0$) and where
$\omega_\mathbf{q}^\mathrm{ph}=c|\mathbf{q}|$ is the photon dispersion with $c$ the speed of light
\cite{cohen_QED}. 
Notice that in \eqref{eq:H_ph},
the summation over photon polarizations
has been made implicit in order to simplify the notation in the sequel of the paper.

In the long-wavelength limit (dipolar approximation), the localized surface plasmons interact with the light modes through the Hamiltonian 
\begin{equation}
\label{eq:H_pl-ph}
H_{\mathrm{pl}\textrm{-}\mathrm{ph}}=\sum_\mathbf{R}\left[\frac{Q}{M}\boldsymbol{\Pi}(\mathbf{R})\cdot\mathbf{A}(\mathbf{R})
+\frac{Q^2}{2M}\mathbf{A}^2(\mathbf{R})\right],
\end{equation}
where 
\begin{equation}
\label{eq:A}
\mathbf{A}(\mathbf{R})=\sum_{\mathbf{q}}
\hat{\boldsymbol{\epsilon}}\sqrt{\frac{\hbar}{2\epsilon_0\mathcal{V}\omega_\mathbf{q}^\mathrm{ph}}}
\left(c_\mathbf{q}\, \mathrm{e}^{\mathrm{i}\mathbf{q}\cdot\mathbf{R}}
+c_\mathbf{q}^\dagger\,\mathrm{e}^{-\mathrm{i}\mathbf{q}\cdot\mathbf{R}}\right)
\end{equation}
is the vector potential at the location
$\mathbf{R}$ of the nanoparticles. 
Expressing the conjugate momentum 
$\boldsymbol{\Pi}(\mathbf{R})=\Pi(\mathbf{R})\hat{\mathbf{p}}$
in \eqref{eq:A} in terms of the
creation and annihilation operators
associated with localized surface plasmons (see \eqref{eq:b}),
going to Fourier space and using
$\hat{\mathbf{p}}\cdot\hat{\boldsymbol{\epsilon}}=1$, we obtain for \eqref{eq:H_pl-ph}
the expression
\begin{align}
\label{eq:H_pl-ph_1}
H_{\mathrm{pl}\textrm{-}\mathrm{ph}}=&\;
\hbar\omega_0\sum_\q
\Big[
\mathrm{i}\xi_\q\left(b_\q^\dagger c_\q^{\phantom{\dagger}}-c_\q^\dagger b_\q^{\phantom{\dagger}}
+b_\q^\dagger c_{-\q}^\dagger-c_{-\q}^{}b_\q^{}\right)
\nonumber\\
&+\xi_\q^2\left(c_\q^\dagger
c_\q^{\phantom{\dagger}}+c_\q^{\phantom{\dagger}}c_\q^\dagger+c_\q^\dagger
c_{-\q}^\dagger+c_{-\q}^{}c_\q^{} \right)
\Big],
\end{align}
with $\xi_\q=(\pi\omega_0/\omega_\q^\mathrm{ph})^{1/2}(r/a)^{3/2}$.
In terms of the Bogoliubov modes \eqref{eq:beta}
diagonalizing the purely plasmonic problem, 
and using the inverse transform $b_\q=\cosh{\vartheta_\q}\beta_\q+\sinh{\vartheta_\q}\beta_{-\q}^\dagger$, 
equation
\eqref{eq:H_pl-ph_1} thus reads
\begin{align}
\label{eq:H_pl-ph_beta}
H_{\mathrm{pl}\textrm{-}\mathrm{ph}}=&\;\hbar\omega_0\sum_\q
\Big[
\mathrm{i}\xi_\q(\cosh{\vartheta_\q}-\sinh{\vartheta_\q})
\nonumber\\
&\times
\left(\beta_\q^\dagger
c_\q^{\phantom{\dagger}}-c_\q^\dagger \beta_\q^{\phantom{\dagger}}
+\beta_\q^\dagger c_{-\q}^\dagger-c_{-\q}\beta_\q\right)
\nonumber\\
&+\xi_\q^2\left(c_\q^\dagger
c_\q^{\phantom{\dagger}}+c_\q^{\phantom{\dagger}}c_\q^\dagger+c_\q^\dagger
c_{-\q}^\dagger+c_{-\q}c_\q \right)
\Big].
\end{align}

%===========================================================================
%===========================================================================
%===========================================================================
%===========================================================================
\section{Plasmon polariton dispersion}
\label{sec:PP}
The total Hamiltonian
\begin{equation}
\label{eq:Htot}
H=H_\mathrm{pl}+H_\mathrm{ph}+H_{\mathrm{pl}\textrm{-}\mathrm{ph}} 
\end{equation}
with $H_\mathrm{pl}$, $H_\mathrm{ph}$ and $H_{\mathrm{pl}\textrm{-}\mathrm{ph}}$
given, respectively, in \eqref{eq:H_pl_diag}, \eqref{eq:H_ph} and
\eqref{eq:H_pl-ph_beta}, can now be diagonalized by introducing the annihilation operator
associated with plasmon polaritons 
\begin{equation}
\label{eq:gamma}
\gamma_\q=w_\q^{} c_\q^{}+x_\q^{} \beta_\q^{}+y_\q^{} c_{-\q}^\dagger+z_\q^{} \beta_{-\q}^\dagger, 
\end{equation}
with $w_\q$, $x_\q$, $y_\q$ and $z_\q$ complex numbers. 
As detailed in Appendix \ref{sec:PPdisp},
imposing the diagonal form of the Heisenberg equation of motion $[\gamma_\q, H]=\hbar\omega_\q^\mathrm{PP}\gamma_\q$
yields the 
frequency- and wavevector-dependent dielectric function 
\begin{equation}
\label{diel}
\epsilon(\q,
\omega)=\frac{c^2\q^2}{\omega^2}=1+\frac{8\pi\Omega\omega_0}{{\omega_\q^\mathrm{pl}}^2-\omega^2}.
\end{equation}
The spatial dispersion of the dielectric function (i.e., its dependence on $\q$) stems from the 
interaction between LSPs leading to the plasmonic dispersion $\omega_\q^\mathrm{pl}$ in our system.
This is reminiscent of the case of exciton polaritons in bulk semiconductors \cite{agranovich}. A crucial difference of our system, however, is 
that the dielectric function of the metamaterial in (\ref{diel}) is strongly sensitive to the polarization of
incoming light, which quantitatively affects the collective plasmon dispersion $\omega_\q^\mathrm{pl}$.
This effect is a direct consequence of the 
anisotropic dipolar interaction between the metallic nanoparticles, resulting in a birefringence of
the plasmonic metamaterial despite the  symmetric lattice structure of our array. This is in stark
contrast with conventional birefringence observed in crystals, which is usually associated with
strongly asymmetric lattice structures \cite{born}.

\begin{figure}[tb]
\centerline{\includegraphics[width=\columnwidth]{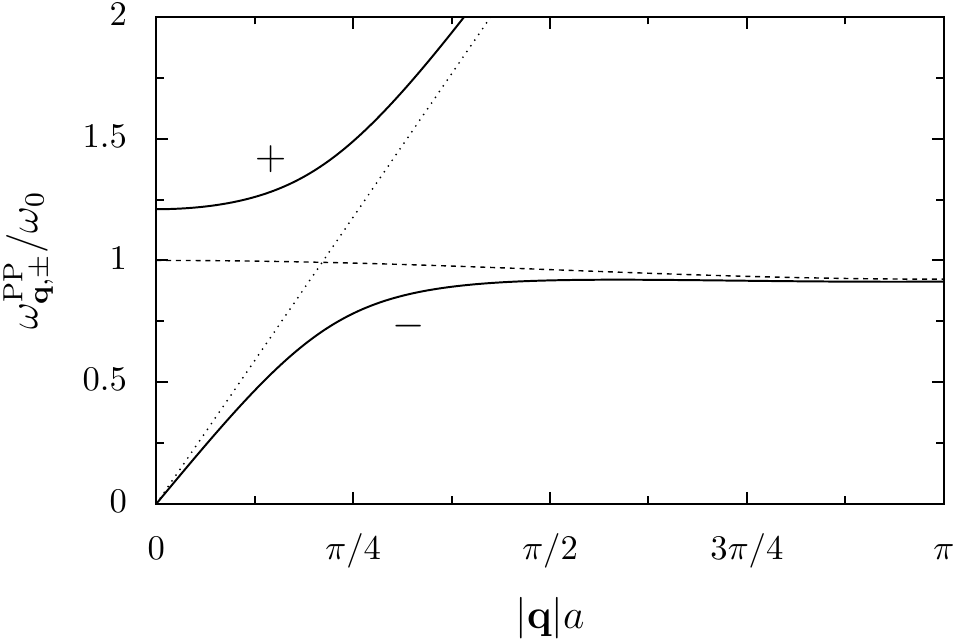}}
\caption{\label{fig:PP_100}%
Solid lines: plasmon polariton dispersion \eqref{eq:PP} along the $\Gamma\mathrm{X}$
direction in the first Brillouin zone (see the inset in Fig.\
\ref{fig:plasmon_dispersion})
and for transverse light polarization
($\hat{\mathbf{q}}\cdot\hat{\boldsymbol{\epsilon}}=0$).
Dotted line: dispersion $\omega_\q^\mathrm{ph}$ of free light.
Dashed line: collective plasmon dispersion
\eqref{eq:omega_pl}.
In the figure, $a=3r$ and $c/\omega_0=3a/2$.}
\end{figure}

The dielectric function \eqref{diel} 
results in the plasmon polariton dispersion
\begin{align}
\label{eq:PP}
\omega_{\q, \pm}^\mathrm{PP}=&\; \frac{1}{\sqrt{2}}
\Bigg[
{\omega_\q^\mathrm{pl}}^2+{\omega_\q^\mathrm{ph}}^2+8\pi\Omega\omega_0
\nonumber\\
&\pm\sqrt{\left({\omega_\q^\mathrm{pl}}^2+{\omega_\q^\mathrm{ph}}^2+8\pi\Omega\omega_0\right)^2
-4{\omega_\q^\mathrm{pl}}^2{\omega_\q^\mathrm{ph}}^2}
\Bigg]^{1/2}
\end{align}
which is shown in Fig.\
\ref{fig:PP_100} along the $\Gamma\mathrm{X}$ direction in the first Brillouin zone 
for transverse light
polarization (i.e., in the $yz$ plane, see the inset in Fig.\ \ref{fig:plasmon_dispersion}).
For wavenumbers close to the edge of the first Brillouin zone, the $+$ and $-$ branches of the
plasmon polariton dispersion \eqref{eq:PP} asymptotically approach 
the light and the collective plasmon dispersion,
respectively. When $|\q|\to 0$, however, the $-$ branch goes to $\omega_\q^\mathrm{ph}\to0$, while the
$+$ branch tends to $\omega_{|\q|\to0,
+}^\mathrm{PP}\simeq\omega_0(1+8\pi\Omega/\omega_0)^{1/2}$. 
Thus the strong plasmon-photon coupling results in a gap of the order of  $\Delta\simeq4\pi\Omega$
in the plasmon polariton dispersion.
This has important consequences on the optical properties of
our simple cubic array of nanoparticles. Indeed, for frequencies within the band
gap, no plasmon polariton can propagate in
the metamaterial, such that the reflectivity of the latter is equal to one. 
We estimate that 
for an interparticle distance $a=3r$, the polaritonic gap $\Delta$ is 
about $\unit[25]{\%}$ of the Mie frequency $\omega_0$.
For noble-metal nanoparticles, the latter usually lies in the visible range
($\omega_0\simeq\unit[2-3]{eV/\hbar}$), yielding a polaritonic gap of about
$\Delta\simeq\unit[0.5-0.75]{eV/\hbar}$.

\begin{figure}[tb]
\centerline{\includegraphics[width=\columnwidth]{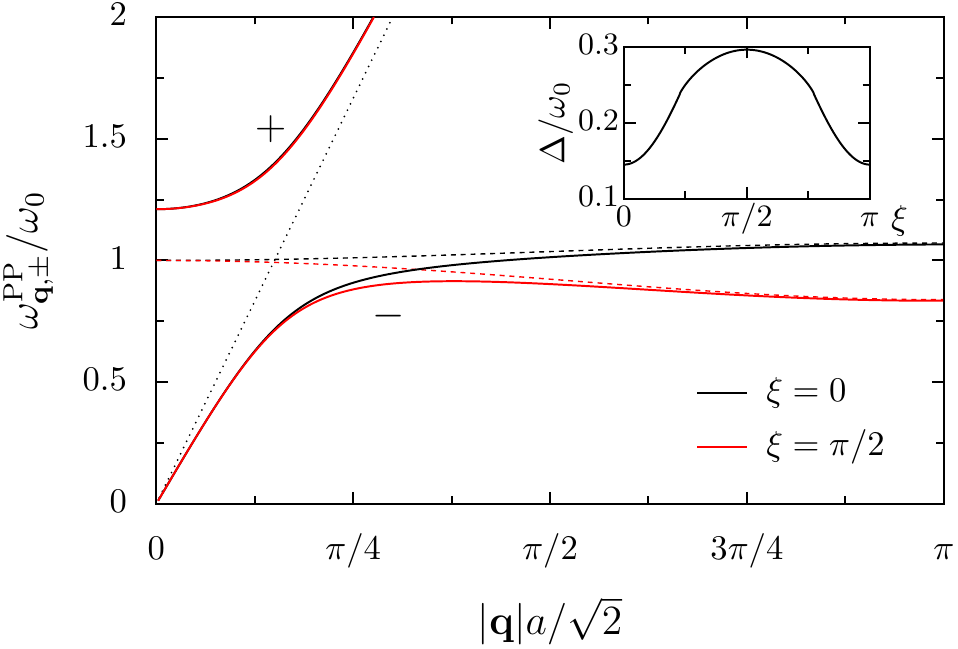}}
\caption{\label{fig:PP_110}%
Solid lines: plasmon polariton dispersion \eqref{eq:PP} along the $\Gamma\mathrm{M}$
direction in the first Brillouin zone 
and for two transverse light polarizations characterized by the angle $\xi$ (see text).
Dotted line: dispersion $\omega_\q^\mathrm{ph}$ of light.
Dashed lines: collective plasmon dispersion \eqref{eq:omega_pl}.
Inset: polaritonic band gap as a function of polarization angle.
Same parameters as in Fig.\ \ref{fig:PP_100}.}
\end{figure}

Remarkably, the plasmon polariton dispersion \eqref{eq:PP} can be tuned by the polarization of light
through the modification of the collective plasmon dispersion
\eqref{eq:omega_pl} (see 
Fig.~\ref{fig:plasmon_dispersion}). This is illustrated
in Fig.\ \ref{fig:PP_110}, which shows the plasmon
polariton dispersion along the $\Gamma\mathrm{M}$ direction in the first Brillouin
zone (see the inset in Fig.\ \ref{fig:plasmon_dispersion}) for two polarization angles $\xi$ defined through
the transverse polarization
$\hat{\boldsymbol{\epsilon}}=\cos{\xi}\;\hat{\boldsymbol{\epsilon}}_1+\sin{\xi}\;\hat{\boldsymbol{\epsilon}}_2$,
with
$\hat{\boldsymbol{\epsilon}}_1=\hat{\mathbf{z}}\times\hat{\mathbf{q}}/|\hat{\mathbf{z}}\times\hat{\mathbf{q}}|$
and 
$\hat{\boldsymbol{\epsilon}}_2=\hat{\mathbf{q}}\times\hat{\boldsymbol{\epsilon}}_1/|\hat{\mathbf{q}}\times\hat{\boldsymbol{\epsilon}}_1|$.
As can be seen from Fig.\ \ref{fig:PP_110}, 
the $-$ branch of the plasmon polariton is significantly modulated by the
polarization of light. This effect results from the dependence on polarization of the collective plasmon
dispersion (dashed lines in the figure) and on the consequent spatial dispersion of the dielectric function \eqref{diel}.\footnote{Within our nearest-neighbour coupling approximation, there are, however, exceptional directions in $q$-space for which the plasmon
polariton band structure does not depend on the direction of the transverse
polarization for symmetry reasons, e.g., the $\Gamma\mathrm{X}$ and $\Gamma\mathrm{R}$ directions.} Hence, the polaritonic band gap, defined
as $\Delta=\omega_{|\q|=0,+}^\mathrm{PP}-\max(\omega_{\q, -}^\mathrm{PP})$, can be significantly
modulated (by about $\unit[50]{\%}$) by tilting the polarization of light, as
shown in the inset of Fig.\ \ref{fig:PP_110}. Considering the amplitude of the effect, this 
feature should be clearly measurable in an experiment measuring the reflectivity of the metamaterial as a function of frequency. The size of the stop band presenting perfect reflection should thus be significantly modulated by the polarization of incoming light.

Plasmonic damping, that may mask the above gap and its modulation, is of crucial importance for the experimental
observability of the effect we are predicting. The plasmon polaritons are mainly subject to two sources of dissipation:
Ohmic (absorption) losses and Landau damping, i.e., the decay into electron-hole pairs
\cite{kawab66_JPSJ, weick05_PRB}. In the present context, radiation damping is irrelevant 
due to the very definition of plasmon polaritons which are eigenstates of the coupled
plasmon-photon system. A conservative estimate of the Ohmic losses from a Drude model yields, at
room temperature and for silver, a Drude linewidth $\gamma_\mathrm{D}\simeq\unit[17]{meV}/\hbar$
\cite{ashcroft}. The Landau damping decay rate can be expressed for an isolated metallic
nanoparticle as $\gamma_\mathrm{L}=3v_\mathrm{F}g/4r$, 
where $v_\mathrm{F}$ is the Fermi velocity and $g$ a numerical factor 
of order one \cite{kawab66_JPSJ}.\footnote{This estimate still holds for near-field 
coupled nanoparticles, see \cite{adam}.}
For Ag nanoparticles, one has $\hbar\gamma_\mathrm{L}\simeq \unit[690]{meV}/r[\mathrm{nm}]$, so that 
for the nanoparticle sizes we consider, typically with radii of the order of
$\unit[10]{nm}$, the total plasmonic linewidth is up to 
$\unit[0.1]{eV}/\hbar$ for silver nanoparticles. As a consequence, 
the polaritonic band gap and its modulation estimated above should be clearly
observable in an experiment.
Our estimate of the plasmon polariton linewidth also shows that it is
dominated by Landau damping. 
The quantum origin of this dissipative mechanism and its predominance with
respect to absorption losses further justifies  
our quantum approach.\footnote{
For this reason, we also do not take into account absorption losses when
quantizing the radiation in the medium. See, e.g., \cite{vogel}.}

%===========================================================================
%===========================================================================
%===========================================================================
%===========================================================================
\section{Conclusion}
\label{sec:ccl}
We have developed an analytical quantum theory of the
strong coupling regime between photons and collective plasmons in three-dimensional
arrays of interacting metallic nanoparticles. 
Remarkably, the resulting plasmon polaritons present a band structure that can be significantly
modulated by the polarization of light. Such a tunability crucially stems from the dipolar
interactions between the localized surface plasmons in each nanoparticle.
As a result, the dielectric function and thus the reflection and transmission
coefficients of the metamaterial can be tuned by changing the polarization of 
light. The consequent optical birefringence is directly due to the anisotropic dipolar 
interactions between nanoparticles despite the symmetric lattice structure of the metamaterial. 

Our results obtained for a simple cubic array can be easily extended 
to other types of metastructures,
such as bcc, fcc, or hcp lattices of metallic nanoparticles, paving the way to quantum
plasmonic metamaterials with fully tunable optical properties.

%===========================================================================
%===========================================================================
%===========================================================================
%===========================================================================
\begin{acknowledgement} 
We thank W.\ L.\ Barnes, R.\ Caroni, B.\ Donnio, S.\
Foteinopoulou, J.-L.\ Gallani, 
P.\ Gilliot, O.\ Hess, R.\ A.\ Jalabert and D.\ Weinmann for valuable discussions and useful comments.
We acknowledge the CNRS PICS program (Contract No.\ 6384 APAG), the French
National Research Agency ANR (Project No.\ ANR-14-CE26-0005-01 Q-MetaMat), and
the Royal Society (International Exchange Grant No.\ IE140367) for financial support.
\end{acknowledgement}

%===========================================================================
%===========================================================================
%===========================================================================
%===========================================================================
\appendix
%===========================================================================
%===========================================================================
%===========================================================================
%===========================================================================
\section{Collective plasmon dispersion with dipole-dipole interaction beyond nearest neighbors}
\label{sec:beyond}
For simplicity of treatment and to highlight the main physical concepts behind
our work, in Sect.\ \ref{sec:plasmon} we only discuss the effects of interactions between nearest neighboring nanoparticles in the simple cubic lattice. However, as the dipole-dipole interaction decays as one
over the cube of the interparticle distance, it is important to check the robustness of our
results against the effect of interactions beyond nearest neighbors. In the
following, we show that the plasmon dispersion discussed in Sect.\
\ref{sec:plasmon} is not qualitatively 
modified by interactions beyond nearest neighbors. Specifically, we analytically compute the collective plasmon
dispersion including next, third and fourth nearest neighbors and show that the interaction between the
nearest neighbors alone captures the relevant physics of the problem.

The purely plasmonic Hamiltonian
\eqref{eq:H_pl_0} has to be supplemented by three extra terms 
$H_\mathrm{int}^{(2)}$, $H_\mathrm{int}^{(3)}$ and $H_\mathrm{int}^{(4)}$
when the dipole-dipole interaction between next, third and fourth nearest neighbors are taken into
account, viz.\
\begin{equation}
\label{eq:H_pl_99}
H_\mathrm{pl}=H_0+H_\mathrm{int}+\sum_{n=2}^4H_\mathrm{int}^{(n)},
\end{equation}
with
\begin{subequations}
\begin{align}
H_\mathrm{int}^{(2)}=&\;\frac{Q^2}{8\pi\epsilon_0(\sqrt{2}a)^3}
\sum_{\mathbf{R}}\sum_{j=1}^3\sum_{\sigma=\pm}
\mathcal{C}_{j\sigma}^{(2)}h(\mathbf{R})
\nonumber\\
&\times
\big[h(\mathbf{R}+\mathbf{e}_{j\sigma}^{(2)})+h(\mathbf{R}-\mathbf{e}_{j\sigma}^{(2)})\big], 
\\
H_\mathrm{int}^{(3)}=&\;\frac{Q^2}{8\pi\epsilon_0(\sqrt{3}a)^3}
\sum_{\mathbf{R}}\sum_{j=1}^2\sum_{\sigma=\pm}
\mathcal{C}_{j\sigma}^{(3)}h(\mathbf{R})
\nonumber\\
&\times
\big[h(\mathbf{R}+\mathbf{e}_{j\sigma}^{(3)})+h(\mathbf{R}-\mathbf{e}_{j\sigma}^{(3)})\big], 
\\
H_\mathrm{int}^{(4)}=&\;\frac{Q^2}{8\pi\epsilon_0(2a)^3}
\sum_{\mathbf{R}}\sum_{j=1}^3
\mathcal{C}_{j}h(\mathbf{R})
\nonumber\\
&\times
\big[h(\mathbf{R}+2\mathbf{e}_{j})+h(\mathbf{R}-2\mathbf{e}_{j})\big].
\end{align}
\end{subequations}
Here, $H_\mathrm{int}^{(n)}$ represents
the dipole-dipole interaction Hamiltonian between the $n$th nearest neighbors.
We define 
\begin{subequations}
\begin{align}
\mathcal{C}_{1\sigma}^{(2)}&=1-\frac32\sin^2{\theta}\left(\cos{\varphi}+\sigma\sin{\varphi}\right)^2,\\
\mathcal{C}_{2\sigma}^{(2)}&=1-\frac32\left(\sin{\theta}\sin{\varphi}+\sigma\cos{\theta}\right)^2,\\
\mathcal{C}_{3\sigma}^{(2)}&=1-\frac32\left(\sin{\theta}\cos{\varphi}+\sigma\cos{\theta}\right)^2,\\
\mathcal{C}_{1\sigma}^{(3)}&=1-\left(\sin{\theta}\cos{\varphi}+\sin{\theta}\sin{\varphi}+\sigma\cos{\theta}\right)^2,\\
\mathcal{C}_{2\sigma}^{(3)}&=1-\left(\sin{\theta}\cos{\varphi}-\sin{\theta}\sin{\varphi}+\sigma\cos{\theta}\right)^2,
\end{align}
\end{subequations}
as well as the vectors
$\mathbf{e}_{1\sigma}^{(2)}=\mathbf{e}_1+\sigma\mathbf{e}_2$,
$\mathbf{e}_{2\sigma}^{(2)}=\mathbf{e}_2+\sigma\mathbf{e}_3$, 
$\mathbf{e}_{3\sigma}^{(2)}=\mathbf{e}_1+\sigma\mathbf{e}_3$, 
$\mathbf{e}_{1\sigma}^{(3)}=\mathbf{e}_1+\mathbf{e}_2+\sigma\mathbf{e}_3,$ and 
$\mathbf{e}_{2\sigma}^{(3)}=\mathbf{e}_1-\mathbf{e}_2+\sigma\mathbf{e}_3$.
Using the bosonic ladder operator in momentum space $b_\mathbf{q}$ and its
adjoint,
the plasmonic Hamiltonian \eqref{eq:H_pl_99} including 
dipole-dipole interactions up to the fourth nearest neighbors then reads
\begin{align}
\label{eq:H_pl_(2)}
H_\mathrm{pl}=&\;\hbar\sum_{\mathbf{q}}
\left\{
\left[\omega_0+2\Omega \left(f_\q+\sum_{n=2}^4f_\q^{(n)}\right)\right] b_\q^\dagger b_\q^{\phantom{\dagger}}
\right.\nonumber\\
&\left.+\ \Omega \left(f_\q+\sum_{n=2}^4f_\q^{(n)}\right)\left(b_\q^\dagger
b_{-\q}^\dagger+b_{-\q}b_{\q}\right)
\right\}, 
\end{align}
with 
\begin{subequations}
\begin{align}
f_\q^{(2)}&=\frac{1}{2\sqrt{2}}\sum_{j=1}^3\sum_{\sigma=\pm}\mathcal{C}_{j\sigma}^{(2)}
\cos{(\q\cdot\mathbf{e}_{j\sigma}^{(2)})}, \\
f_\q^{(3)}&=\frac{1}{3\sqrt{3}}\sum_{j=1}^2\sum_{\sigma=\pm}\mathcal{C}_{j\sigma}^{(3)}
\cos{(\q\cdot\mathbf{e}_{j\sigma}^{(3)})}, \\
f_\q^{(4)}&=\frac{1}{8}\sum_{j=1}^3\mathcal{C}_{j}
\cos{(2\q\cdot\mathbf{e}_{j})}.
\end{align}
\end{subequations}

\begin{figure*}[tb]
\includegraphics[width=\linewidth]{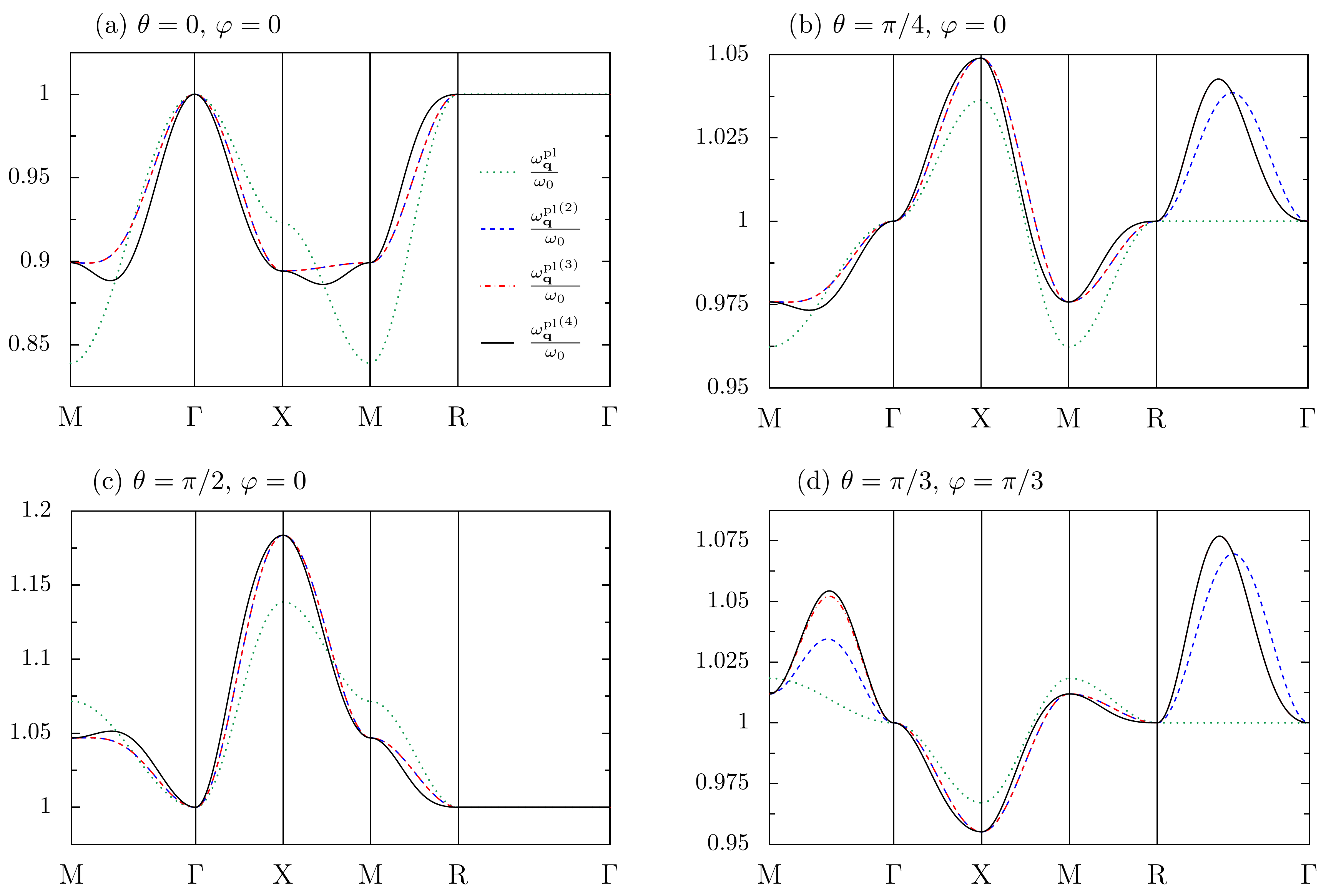}
\caption{\label{fig:plasmon_dispersion_fourth_nn}%
Collective plasmon dispersion relation with dipole-dipole interaction including
nearest (dotted), next nearest (dashed), third nearest (dashed-dotted) and
fourth nearest neighbors (solid lines) along the
high symmetry axes in the first Brillouin zone, for various
polarization angles $(\theta, \varphi)$. In the figure, $r=a/3$.}
\end{figure*}

As explicited in Sect.\ \ref{sec:plasmon}, the Hamiltonian \eqref{eq:H_pl_(2)} above can be
diagonalized by a Bogoliubov transformation, leading to the collective plasmon
dispersion
\begin{equation}
\label{eq:omega_pl_(4)}
{\omega_\q^\mathrm{pl}}^{(4)}=\omega_0\sqrt{1+4\frac{\Omega}{\omega_0}
\left(f_\q+\sum_{n=2}^4f_\q^{(n)}\right)}.
\end{equation}
The dispersion relation \eqref{eq:omega_pl_(4)} is shown in Fig.\
\ref{fig:plasmon_dispersion_fourth_nn} for various polarization angles of the
localized surface plasmons (solid lines). For comparison, we also show in Fig.\ \ref{fig:plasmon_dispersion_fourth_nn} 
the plasmon dispersion \eqref{eq:omega_pl}
that only includes nearest
neighbor interactions (dotted lines), as well as the dispersion
\begin{equation}
{\omega_\q^\mathrm{pl}}^{(2)}=\omega_0\sqrt{1+4\frac{\Omega}{\omega_0}
\left(f_\q+f_\q^{(2)}\right)}
\end{equation}
and
\begin{equation}
{\omega_\q^\mathrm{pl}}^{(3)}=\omega_0\sqrt{1+4\frac{\Omega}{\omega_0}
\left(f_\q+\sum_{n=2}^3f_\q^{(n)}\right)}
\end{equation}
that include
the next nearest and third nearest neighbors (dashed and dashed-dotted lines in
Fig.\ \ref{fig:plasmon_dispersion_fourth_nn}, respectively).
As can be seen from the figure, dipole-dipole interactions beyond the 
nearest neighbor contribution do not lead to dramatic qualitative changes in the collective plasmon
dispersion relation. For clarity, in the main text we thus limit ourselves to the discussion of dipole-dipole
interaction effects between nearest neighbors only.

%===========================================================================
%===========================================================================
%===========================================================================
%===========================================================================
\section{Derivation of the plasmon polariton dispersion}
\label{sec:PPdisp}
Below, we provide a detailed derivation of the plasmon polariton dispersion
relation \eqref{eq:PP}, as well as of
the associated dielectric function \eqref{diel}, following \cite{hopfi58_PR}.
In terms of the operators $\beta_\q^{}$ and $\beta_\q^\dagger$ ($c_\q^{}$ and $c_\q^\dagger$) annihilating and creating a collective
plasmon (a photon) of momentum $\q$ with dispersion $\omega_\q^\mathrm{pl}$
($\omega_\q^\mathrm{ph}$), respectively, the total Hamiltonian \eqref{eq:Htot} of our system reads 
\begin{align}
\label{eqS:H}
H=&\;\hbar\sum_\q 
\Big[
A_\q^{} c_\q^\dagger c_\q^{}+B_\q^{} \beta_\q^\dagger \beta_\q^{}
\nonumber\\
&+\mathrm{i}C_\q^{} \left(\beta_\q^\dagger
c_\q^{}-c_\q^\dagger \beta_\q^{}
+\beta_\q^\dagger c_{-\q}^\dagger-c_{-\q}^{}\beta_\q^{}\right)
\nonumber\\
&+\ D_\q^{}
\left(c_\q^\dagger
c_\q^{}+c_\q^{}c_\q^\dagger+c_\q^\dagger
c_{-\q}^\dagger+c_{-\q}^{}c_\q^{} \right)\Big]. 
\end{align}
Here we introduced the notation 
\begin{subequations}
\label{eq:AA}
\begin{align}
A_\q&=\omega_\q^\mathrm{ph},\\
\label{eq:B}
B_\q&=\omega_\q^\mathrm{pl},\\
\label{eq:C}
C_\q&=\omega_0\xi_\q(\cosh{\vartheta_\q}-\sinh{\vartheta_\q}), \\
\label{eq:D}
D_\q&=\omega_0\xi_\q^2. 
\end{align}
\end{subequations}
Note that these four coefficients are invariant under the transformation
$\q\rightarrow -\q$. 

Introducing the new bosonic operators \eqref{eq:gamma}
and imposing the diagonal 
form of the Heisenberg 
equation of motion $[\gamma_\q, H]=\hbar \omega\gamma_\q$, we get with \eqref{eqS:H} 
the set of equations
\begin{align}
&\left(\begin{array}{cccc}
A_\q+2 D_\q-\omega & \mathrm{i}C_\q & -2D_\q & \mathrm{i} C_\q \\
-\mathrm{i}C_\q & B_\q-\omega & \mathrm{i}C_\q & 0 \\
2D_\q & \mathrm{i}C_\q & -A_\q-2D_\q-\omega & \mathrm{i}C_\q \\
\mathrm{i}C_\q & 0 & -\mathrm{i}C_\q & -B_\q-\omega
\end{array}
\right)
\nonumber\\
&\times
\left(
\begin{array}{c}
w_\q \\
x_\q \\
y_\q \\
z_\q
\end{array}
\right)
=0.
\end{align}
The system above only has nontrivial solutions when its determinant is zero. Noticing that $B_\q
D_\q-C_\q^2=0$ (see \eqref{eq:AA}) the condition above yields the eigenvalue equation
\begin{equation}
\label{eqS:eigenvalue}
\epsilon(\q, \omega)\equiv \frac{A_\q^2}{\omega^2}=1+\frac{4A_\q D_\q}{B_\q^2-\omega^2},
\end{equation} 
where $\epsilon(\q, \omega)$ is the frequency- and wavevector-dependent
dielectric function of the system. Solving for 
\eqref{eqS:eigenvalue} gives the plasmon polariton dispersion 
\begin{align}
\label{eqS:omega_PP}
\omega_{\q, \pm}^\mathrm{PP}=&\;\frac{1}{\sqrt{2}}
\Big[A_\q^2+B_\q^2+4A_\q D_\q
\nonumber\\
&\pm\sqrt{\left(A_\q^2+B_\q^2+4A_\q D_\q\right)^2-4A_\q^2B_\q^2}
\Big]^{1/2}.
\end{align}
With the coefficients \eqref{eq:AA}, \eqref{eqS:eigenvalue} and \eqref{eqS:omega_PP} then translate
into the dielectric function \eqref{diel} and the plasmon polariton dispersion
\eqref{eq:PP}.

%===========================================================================
%===========================================================================
%===========================================================================
%===========================================================================


\begin{thebibliography}{}

\bibitem{barne03_Nature}
W. L. Barnes, A. Dereux, T. W. Ebbesen, 
\href{http://dx.doi.org/10.1038/nature01937}
{Nature \textbf{424}, 824 (2003)}

\bibitem{maier}
S. A. Maier, 
\textit{Plasmonics: Fundamentals and Applications} 
(Springer-Verlag, Berlin, 2007)

\bibitem{pendry12_Science}
J. B. Pendry, A. Aubry, D. R. Smith, S. A. Maier, 
\href{http://dx.doi.org/10.1126/science.1220600}
{Science \textbf{337}, 549 (2012)}

\bibitem{vesel68_SPU}
V. G. Veselago, 
\href{http://dx.doi.org/10.1070/PU1968v010n04ABEH003699}
{Sov. Phys. Usp. \textbf{10}, 509 (1968)}

\bibitem{smith00_PRL}
D. R. Smith, W. J. Padilla, D. C. Vier, S. C. Nemat-Nasser, S. Schultz, 
\href{http://dx.doi.org/10.1103/PhysRevLett.84.4184}
{Phys. Rev. Lett. \textbf{84}, 4184 (2000)}

\bibitem{shelb01_Science}
R. A. Shelby, D. R. Smith, S. Schultz, 
\href{http://dx.doi.org/10.1126/science.1058847}
{Science \textbf{292}, 77 (2001)}

\bibitem{pendr00_PRL}
J. B. Pendry, 
\href{http://dx.doi.org/10.1103/PhysRevLett.85.3966}
{Phys. Rev. Lett. \textbf{85}, 3966 (2000)}

\bibitem{fang05_Science}
N. Fang, H. Lee, C. Sun, X. Zhang, 
\href{http://dx.doi.org/10.1126/science.1108759}
{Science \textbf{308}, 534 (2005)}

\bibitem{leonh06_Science}
U. Leonhardt, 
\href{http://dx.doi.org/10.1126/science.1126493}
{Science \textbf{312}, 1777 (2006)}

\bibitem{pendr06_Science}
J. B. Pendry, D. Schurig, D. R. Smith, 
\href{http://dx.doi.org/10.1126/science.1125907}
{Science \textbf{312}, 1780 (2006)}

\bibitem{schur06_Science}
D. Schurig, J. J. Mock, B. J. Justice, S. A. Cummer, J. B. Pendry, A. F. Starr, D. R. Smith, 
\href{http://dx.doi.org/10.1126/science.1133628}
{Science \textbf{314}, 977 (2006)}

\bibitem{tsakm07_Nature}
K. L. Tsakmakidis, A. D. Boardman, O. Hess, 
\href{http://dx.doi.org/10.1038/nature06285}
{Nature \textbf{450}, 397 (2007)}

\bibitem{silva14_Science}
A. Silva, F. Monticone, G. Castaldi, V. Galdi, A. Al\`u, N. Engheta, 
\href{http://dx.doi.org/10.1126/science.1242818}
{Science \textbf{343}, 160 (2014)}

\bibitem{tame13_NaturePhys}
M. S. Tame, K. R. McEnery, \c S. K. Özdemir, J. Lee, S. A. Maier, M. S.
Kim, 
\href{http://dx.doi.org/10.1038/nphys2615}
{Nat. Phys. \textbf{9}, 329 (2013)}

\bibitem{krenn99_PRL}
J. R. Krenn, A. Dereux, J. C. Weeber, E. Bourillot, Y. Lacroute, J. P.
Goudonnet, G. Schider, W. Gotschy, A. Leitner, F. R. Aussenegg, C. Girard, 
\href{http://dx.doi.org/10.1103/PhysRevLett.82.2590}
{Phys. Rev. Lett. \textbf{82}, 2590 (1999)}

\bibitem{maier02_PRB}
S. A. Maier, M. L. Brongersma, P. G. Kik, H. A. Atwater, 
\href{http://dx.doi.org/10.1103/PhysRevB.65.193408}
{Phys. Rev. B \textbf{65}, 193408 (2002)}

\bibitem{felid02_PRB}
N. F\'elidj, J. Aubard, G. L\'evi, J. R. Krenn, G. Schider, A. Leitner, F.
R. Aussenegg, 
\href{http://dx.doi.org/10.1103/PhysRevB.66.245407}
{Phys. Rev. B \textbf{66}, 245407 (2002)}

\bibitem{maier03_NatureMat}
S. A. Maier, P. G. Kik, H. A. Atwater, S. Meltzer, E. Harel, B. E. Koel, A.
A. G. Requicha, 
\href{http://dx.doi.org/10.1038/nmat852}
{Nature Mater. \textbf{2}, 229 (2003)}

\bibitem{sweat05_PRB}
L. A. Sweatlock, S. A. Maier, H. A. Atwater, J. J. Penninkhof, A. Polman, 
\href{http://dx.doi.org/10.1103/PhysRevB.71.235408}
{Phys. Rev. B \textbf{71}, 235408 (2005)}

\bibitem{polyu11_NL}
D. K. Polyushkin, E. Hendry, E. K. Stone, W. L. Barnes, 
\href{http://dx.doi.org/10.1021/nl202428g}
{Nano Lett. \textbf{11}, 4718 (2011)}

\bibitem{quint98_OL}
M. Quinten, A. Leitner, J. R. Krenn, F. R. Aussenegg, 
\href{http://dx.doi.org/10.1364/OL.23.001331}
{Opt. Lett. \textbf{23}, 1331 (1998)}

\bibitem{brong00_PRB}
M. L. Brongersma, J. W. Hartman, H. A. Atwater, 
\href{http://dx.doi.org/10.1103/PhysRevB.62.R16356}
{Phys. Rev. B \textbf{62}, R16356 (2000)}

\bibitem{park04_PRB}
S. Y. Park, D. Stroud, 
\href{http://dx.doi.org/10.1103/PhysRevB.69.125418}
{Phys. Rev. B \textbf{69}, 125418 (2004)}

\bibitem{weber04_PRB}
W. H. Weber, G. W. Ford, 
\href{http://dx.doi.org/10.1103/PhysRevB.70.125429}
{Phys. Rev. B \textbf{70}, 125429 (2004)}

\bibitem{koend06_PRB}
A. F. Koenderink, A. Polman, 
\href{http://dx.doi.org/10.1103/PhysRevB.74.033402}
{Phys. Rev. B \textbf{74}, 033402 (2006)}

\bibitem{koend09_NL}
A. F. Koenderink, 
\href{http://dx.doi.org/10.1021/nl902439n}
{Nano Lett. \textbf{9}, 4228 (2009)}

\bibitem{weick13_PRL}
G. Weick, C. Woollacott, W. L. Barnes, O. Hess, E. Mariani, 
\href{http://dx.doi.org/10.1103/PhysRevLett.110.106801}
{Phys. Rev. Lett. \textbf{110}, 106801 (2013)}

\bibitem{fano56_PR}
U. Fano, 
\href{http://dx.doi.org/10.1103/PhysRev.103.1202}
{Phys. Rev. \textbf{103}, 1202 (1956)}

\bibitem{hopfi58_PR}
J. J. Hopfield, 
\href{http://dx.doi.org/10.1103/PhysRev.112.1555}
{Phys. Rev. \textbf{112}, 1555 (1958)}

\bibitem{donni07_AdvMater}
B. Donnio, P. Garc\'ia-V\'azquez, J.-L. Gallani, D. Guillon, E. Terazzi, 
\href{http://dx.doi.org/10.1002/adma.200701252}
{Adv. Mater. \textbf{19}, 3534 (2007)}

\bibitem{tan11_NatureNanotech}
S. J. Tan, M. J. Campolongo, D. Luo, W. Cheng, 
\href{http://dx.doi.org/10.1038/NNANO.2011.49}
{Nature Nanotech.\ \textbf{6}, 268 (2011)}

\bibitem{kreibig}
U. Kreibig, M. Vollmer, 
\textit{Optical Properties of Metal Clusters} 
(Springer-Verlag, Berlin, 1995).

\bibitem{gerch02_PRA}
L. G. Gerchikov, C. Guet, A. N. Ipatov, 
\href{http://dx.doi.org/10.1103/PhysRevA.66.053202}
{Phys. Rev. A \textbf{66}, 053202 (2002)}

\bibitem{weick06_PRB}
G. Weick, G.-L. Ingold, R. A. Jalabert, D. Weinmann, 
\href{http://dx.doi.org/10.1103/PhysRevB.74.165421}
{Phys. Rev. B \textbf{74}, 165421 (2006)}

\bibitem{cohen_QED}
C. Cohen-Tannoudji, J. Dupont-Roc, G. Grynberg, 
\textit{Photons and Atoms: Introduction to Quantum Electrodynamics} 
(Wiley-VCH, Weinheim, 2004)

\bibitem{agranovich}
V. M. Agranovich, V. L. Ginzburg, 
\textit{Crystal Optics with Spatial Dispersion, and Excitons}
(Springer-Verlag, Berlin, 1984)

\bibitem{born}
M. Born, E. Wolf, 
\textit{Principles of Optics} 
(Pergamon Press, Oxford, 1980)

\bibitem{kawab66_JPSJ}
A. Kawabata, R. Kubo,
\href{http://dx.doi.org/10.1143/JPSJ.21.1765}
{J. Phys. Soc. Jpn. \textbf{21}, 1765 (1966)}

\bibitem{weick05_PRB}
G. Weick, R. A. Molina, D. Weinmann, R. A. Jalabert, 
\href{http://dx.doi.org/10.1103/PhysRevB.72.115410}
{Phys. Rev. B \textbf{72}, 115410 (2005)}

\bibitem{ashcroft}
N. W. Ashcroft, N. D. Mermin, 
\textit{Solid State Physics} 
(Hartcourt, Fort Worth, 1976)

\bibitem{adam}
A.\ Brandstetter-Kunc, G.\ Weick, D.\ Weinmann, R.\ A.\ Jalabert, 
\href{http://arxiv.org/abs/1407.6569}
{arXiv:1407.6569} (2014)

\bibitem{vogel}
W. Vogel, D. G. Welsch, 
\textit{Quantum Optics} 
(Wiley-VCH, Weinheim, 2006)

\end{thebibliography}
\end{document}